\documentclass[12pt,prd,draft,nofootinbib]{revtex4}

\usepackage{amsfonts}
\usepackage{amssymb}
\usepackage{amsmath}
\usepackage{bm}
\usepackage{latexsym}

\newcommand{\beq}{\begin{eqnarray}}
\newcommand{\eeq}{\end{eqnarray}}
\newcommand{\be}{\begin{equation}}
\newcommand{\ee}{\end{equation}}
\def\la{\mathrel{\mathpalette\fun <}}

\def\fun#1#2{\lower3.6pt\vbox{\baselineskip0pt\lineskip.9pt
\ialign{$\mathsurround=0pt#1\hfil ##\hfil$\crcr#2\crcr\sim\crcr}}}

\newcommand{{\SD}}{\rm SD}

\newcommand{\vep}{\bm p}

\newcommand{\lan}{\langle}
\newcommand{\ran}{\rangle}
\begin{document}
\title{The running mass $m_s$ at low scale from the heavy-light 
meson decay constants}
\author{A.M.Badalian}
\affiliation{State Research Center,
Institute of Theoretical and Experimental Physics,
Moscow, 117218 Russia} 
\email[E-mail:]{badalian@itep.ru}
\author{B.L.G. Bakker}
\affiliation{Department of Physics and Astronomy,
Vrije Universiteit, Amsterdam}
\email[E-mail:]{blg.bakker@few.vu.nl}
\date{\today}

\begin{abstract}
It is shown that a 25(20)\% difference between the decay constants
 $f_{D_s}(f_{B_s})$ and $f_D(f_B)$ occurs due to large differences
in the pole masses of the $s$ and $d(u)$ quarks. The values $\eta_D
=f_{D_s}/f_D\approx 1.23(15)$, recently observed in the CLEO experiment,
 and $\eta_B=f_{B_s}/f_B\approx 1.20$, obtained in unquenched lattice
QCD, can be reached only if in the relativistic Hamiltonian the running
mass  $m_s$ at low scale is $m_s(\sim 0.5 $ GeV)$= 170 - 200$ MeV. Our
results follow from the analytical expression for the pseudoscalar
decay constant $f_{\rm P}$ based on the path-integral representation
of the meson Green's function.

\end{abstract}

\maketitle

Relativistic potential models (RPM) have been successful in
their description of light-light and heavy-light (HL) meson spectra
\cite{ref1,ref2}.  Still there exists a fundamental problem, which remains
partly unsolved up to now. It concerns the choice of the quark masses in
the kinetic term of the relativistic Hamiltonian, which in different RPMs
vary in a wide range. For example, HL mesons were studied with the use of
the Dirac equation, taking for the light quark mass $m_n (n=u,d)=7$ MeV
in \cite{ref3} and 72 MeV in \cite{ref4}, and in the Salpeter equation
for the strange quark mass the values $m_s=419$ MeV in \cite{ref5} and
$m_s=180$ MeV in \cite{ref6} have been used. However, in contrast to
constituent quark models, where the constituent mass can be considered
as a fitting parameter, a fundamental relativistic Hamiltonian has to
contain only conventional quark masses--the pole masses. These masses are
now well established for heavy and light quarks \cite{ref7}. They have
been used in the QCD string approach giving a good description of meson
spectra \cite{ref6,ref8,ref9}.  However, the strange quark mass $m_s$
is still not determined at low scale.  At present, owing to the QCD sum
rules calculations \cite{ref10} and lattice QCD \cite{ref11}, $m_s$
is well established at a rather large scale: $m_s(\mu=2\, {\rm GeV})
=90\pm10$ MeV, while in the Hamiltonian approach the mass $m_s$ enters
at a lower scale,  which is evidently smaller that the scale $\mu_c
\approx 1.2$ GeV for the $c$ quark. Therefore it is very important to
find physical quantities which are very sensitive to $m_s$ at low scale:
$\mu_s \leq 1$ GeV. In this letter we show that such information can be
extracted from the analysis of the decay constants of HL mesons, namely,
from the ratios $f_{D_s}/f_D$ and  $f_{B_s}/f_B$.

Recently, direct measurements of the leptonic decay constants
in the processes $D(D_s)\to \mu\nu_\mu$ \cite{ref12,ref13},
and $B\to \tau\mu_\tau$ \cite{ref14,ref15} have been reported. In
Refs.~\cite{ref12,ref13} the CLEO collaboration gives $f_{D_s}=274(20)$
MeV and $f_D=222.6(20)$ MeV with $\eta_D=f_{D_s}/f_D=1.23(15)$,
having reached an accuracy much better than in previous experiments
\cite{ref16}. This central value for $\eta_D$ appears to be larger than
in many theoretical predictions which typically lie in the range $1.0 -
1.15$ \cite{ref17,ref18,ref19,ref20}. Therefore, one can expect that
precise measurements of $\eta_D$ and $\eta_{\rm B}$ in the future can
become a very important criterium to distinguish different theoretical
models and  check their accuracy.  In particular, relatively large values
\be
 \eta_D=1.25(3), \quad \eta_B=1.20(3),
\label{eq1}
\ee 
have been obtained recently in lattice (unquenched) calculations
\cite{ref21,ref22} and also in our paper \cite{ref6}. In this letter
we show that

1. The running mass $m_s(\mu_1)$ at a low scale, $\mu_1\approx 0.5$
GeV, can be extracted from the values $\eta_D$ and $\eta_B$, if they
are known with high accuracy, $\la 5\%$.

2. The values $\eta_D$ and $\eta_{\rm B}$, as given in Eq.~(\ref{eq1}), can
be obtained only if the running mass $m_s(\mu_1)$ lies inside the range
$170 - 200$ MeV. In particular, in the chiral limit, $m_d=m_u=0$, as well
as for $m_d=8$ MeV, and for $m_s(\mu_1) =180$ MeV, the ratios $\eta_D$
and $\eta_B$ calculated here are
\be
 \eta_D=1.25, \quad \eta_B =1.19\,. 
\label{eq2}
\ee

3. The value $m_s(0.5$ GeV) satisfies the relation 
\be
 \frac{m_s(0.5~{\rm GeV})}{m_s(2~{\rm GeV})}\approx 1.97 \, .
\label{eq3}
\ee
In our analysis we use the analytical expression for the leptonic decay
constant in the pseudoscalar (P) channel, derived in Ref.~\cite{ref6}
with the use of the path-integral representation for the correlator
$G_{\rm P}$ of the currents $j_{\rm P}(x)$: 
$G_{\rm P}(x) =\lan j_{\rm P}(x) j_{\rm P}(0)\ran_{\rm vac}$:
\be 
 J_{\rm P}=\int G_{\rm P} (x) d^3 x = 2 N_c \sum_n
 \frac{\lan Y_{\rm P}\ran_n |\varphi_n(0)|^2}{\omega_{qn}
 \omega_{Qn}}e^{-M_n T},
\label{eq4}
\ee
where $M_n$ and $\varphi_n(r)$ are the eigenvalues and eigenfunctions
of the relativistic string Hamiltonian \cite{ref23,ref24,ref25}, while
$\omega_{qn}(\omega_{Qn})$ is the average kinetic energy of a quark
$q(Q)$ for a given $nS$ state:
\be
 \omega_{qn} =\lan \sqrt{m^2_q+ \vep^2}\ran_{nS}, \quad
 \omega_{Qn} =\lan \sqrt{m^2_Q+\vep^2}\ran_{nS}.
\label{eq5}
\ee 
In Eq.~(\ref{eq5}), $m_q(m_Q)$ is the pole mass of the lighter (heavier)
quark in a heavy-light meson. The matrix element $\lan
Y_{\rm P}\ran_n$ refers to the P channel (with exception of the $\pi$
and $K$ mesons where additional chiral terms occur) and was calculated
in Ref.~\cite{ref6},
\be
 \lan Y_{\rm P}\ran_n=m_qm_Q +\omega_{qn} \omega_{Qn}
 -\lan \vep^2\ran_{nS}.
\label{eq6}
\ee 
On the other hand, for the integral $J_{\rm P}$ (\ref{eq4}) one can use
the conventional spectral decomposition:
\be
 J_{\rm P}=\int G_{\rm P} (x) d^3 x =\sum_n \frac{1}{2M_n}
 (f^n_{\rm P})^2 e^{-M_n T}.
\label{eq7}
\ee 
Then from Eqs.~(\ref{eq4}) and (\ref{eq7}) one obtains that 
\be
 (f^n_{\rm P})^2= \frac{2N_c\lan Y_{\rm P}\ran_n |\varphi_n(0)|^2}{\omega_{qn}
 \omega_{Qn} M_n}.
\label{eq8}
\ee 
All necessary factors in Eq.~(\ref{eq8}) for the ground state $(n=1)$ and
the first radial excitation $(n=2)$ are calculated in Ref.~\cite{ref6} but
here we consider only ground states and  omit the index $n$ everywhere.
Our calculations are performed with the static potential $V_0(r)=\sigma
r -\frac43 \frac{\alpha_{\rm B}(r)}{r}$ \cite{ref9}, and the hyperfine
and self-energy contributions are considered as a perturbation. It is
important that our input parameters contain only fundamental values:
the string tension $\sigma$, the QCD constant $\Lambda (n_f=3)$ in
$\alpha_{\rm B}(r)$, and the conventional pole quark masses.  For the
$s$-quark mass $m_s(\mu_1)$ one can expect that the scale $\mu_1$ is close
to the characteristic momentum $\mu_1 \approx \sqrt{\lan \vep^2\ran}
\sim 0.5 - 0.6$ GeV. This scale also corresponds to the r.m.s. radii
$R_M(1S)$ of the meson we consider. For HL mesons
\be
 R_D\approx R_{D_s} =0.55(1)~{\rm fm},
 \quad R_ B\approx R_{B_s} =0.50(1)~{\rm fm},
\label{eq9}
\ee 
so that  $\mu_1 \sim R^{-1}_M\approx 0.4 - 0.5$ GeV.  We show here that
this mass $m_s(\mu_1)$ is strongly correlated with the values of
$\eta_D$ and $\eta_B$. For other quarks we take $m_c=1.40$ GeV and
$m_b=4.78$ GeV \cite{ref8}.

It is of interest to notice that  for HL mesons the ratios 
\be
 \xi_D =\xi_{D_s} =\frac{|R_{1D}(0)|^2}{\omega_q\omega_c} =0.347(3)
\label{eq10}
\ee
are equal for the $D$ and $D_s$ mesons with an accuracy better than
1\%, and also that these fractions for $B$ and $B_s$ mesons coincide
with $\la 2\%$ accuracy $(\varphi_1^2(0) = R^2_1(0)/4\pi)$:
\be
 \xi_B =\xi_{B_s} =\frac{|R_{1B}(0)|^2}{\omega_q\omega_b} =0.146(2) .
\label{eq11} 
\ee 
It is important that the equalities
$\xi_D=\xi_{D_s}$ and $\xi_B = \xi_{B_s}$ practically do not depend on the
details of the interaction in HL mesons. Therefore, in the ratios
$\eta_D(\eta_B)$ the factors given in Eq.~(\ref{eq10}), $\xi_D(\xi_B)$
cancel and one obtains 
\be 
 \eta^2_{D(B)}= \left(\frac{m_sm_{c(b)}}{\lan Y_{\rm P}\ran_{D(B)}} +
 \frac{\omega_s \omega_{c(b)} - \lan \vep^2\ran_{D_s(B_s)}}
 {\lan Y_{\rm P}\ran_{D(B)}}\right) \frac{M_{D(B)}}{M_{D_s(B_s)}}.  
\label{eq12} 
\ee

In Eq.~(\ref{eq12}) the second term is close to 1.05, while the first term,
proportional to $m_s$, is not small, giving about  30-60\% for
different $m_s$ (below we take $m_s$ from the range $140 \pm 60$ MeV/$c^2$). 
With an accuracy of $\la 2\%$ 
\begin{eqnarray}
 \eta^2_D & =& 2.708\times m_s({\rm GeV})+ 1.07(1), \;
 {\rm if}\; m_d=m_u=0,
\nonumber \\
 \eta^2_{D^+} & =& 2.648\times m_s ({\rm GeV})+ 1.05(1), \;
 {\rm if}\; m_d=8\; {\rm MeV},
\label{eq13}
\end{eqnarray}
i.e., in the chiral limit 
\be
 \eta_D=1.14\;(m_s=85~{\rm MeV}),\quad
 1.25\;(m_s=180~{\rm MeV}),\quad 1.27\;(m_s=200~{\rm MeV}),
\label{eq14}
\ee 
and for $m_d =8$ MeV, $\eta_D =1.13$, 1.24, and 1.26, respectively,
for the same values of $m_s$, so decreasing only by $\sim 1\%$.

For the $B$ and $B_s$ mesons 
\begin{eqnarray}
 \eta^2_B & = &1.90\times m_s + 1.07(1)\quad (m_d=m=0);
\nonumber \\
\eta^2_{B^0}  &= &1.871\times m_s + 1.07(1)\quad (m_d=8~{\rm MeV}),
\label{eq15}
\end{eqnarray}
which practically coincide,  and in the chiral limit $(m_d=m_u=0)$ 
\be
 \eta_{\rm B}=1.11\; (m_s=85~{\rm MeV}),\quad 1.19\;(m_s=180~{\rm MeV})\quad
 1.21\; (m_s=200~{\rm  MeV}).
\label{eq16}
\ee 
These values of $\eta_B$ appear to be only $3 - 5\%$ smaller than
$\eta_D$.

Thus for $m_s=180$ MeV and $m_d=8$ MeV we have obtained  
\be
 \eta_{D^+} =1.25(2), \quad \eta_{B}=1.19(1),
\label{eq17}
\ee 
in good agreement with the CLEO data: $\eta_D({\rm exp})= 1.23(15)$
\cite{ref13}.

To check our choice of $m_s=180$ MeV, we estimate the ratio
$m_s(0.5$ GeV)/$m_s(2$ GeV) using the conventional perturbative
(one-loop) formula for the running mass \cite{ref26} 
\be
 m(\mu^2) =m_0 \left(\frac{1}{2} \ln \frac{\mu^2}{\Lambda^2}\right)^{-d_m}
 \left[1-d_1
\frac{\ln\ln\frac{\mu^2}{\Lambda^2}}{\ln \frac{\mu^2}{\Lambda^2}}\right].
\label{eq18}
\ee 
Here $m_0$ is an integration constant and the other constants are
\begin{equation}
 d_1=\frac{8}{\beta^3_0} \left(51-\frac{19}{3} n_f\right),\; 
 \beta_0 =11 -\frac23 n_f,\;
 d_m =\frac{4}{\beta_0}.
\label{eq19}
\end{equation}
To calculate $m_s$(2 GeV) we take $n_f=4$, $\Lambda(n_f=4) =250$ MeV
$(\beta_0 =\frac{25}{3}$, $d_m =\frac{12}{25}$, $d_1 =0.3548)$. We take
$n_f=3$, $\Lambda(n_f=3)=280$ MeV to determine $m_s$(1 GeV) and
$m_s$(0.5 GeV) (for $n_f =3,~\beta_0=9,~d_m=\frac49,~ d_1=0.3512)$.
Then, from Eqs.~(\ref{eq18},\ref{eq19}) $m_s(2$ GeV) $= 0. 618$ $m_0$,
$m_s$(1 GeV) $= 0.7825~m_0$, and $m_s(0.5 $ GeV) $=1.217~m_0$ and
therefore
\be
 \frac{m_s(1{\rm ~GeV})}{m_s(2 {\rm ~GeV})} =1.27,\quad
\frac{m_s(0.5{\rm ~GeV})}{m_s(2 {\rm ~GeV})} =1.97~.
\label{20}
\ee
It means that $m_s$(0.5 GeV) $=180$ MeV, which we used in our
calculations, corresponds to $m_s($2 GeV) = 91 MeV which coincides with
the conventional value of $m_s$(2 GeV) = 90$\pm15$ MeV \cite{ref7}. Thus
our estimate of $m_s(0.5$ GeV) = 180 MeV supports our choice of this value
in the relativistic string Hamiltonian, which provides a good description
of the HL meson spectra and  decay constants, and gives rise to the
relatively large values of $\eta_D$ and $ \eta_B$ in Eq.~(\ref{eq1}).

\end{document}